# Some Enigmatic Aspects of the Early Universe

C Sivaram, Kenath Arun and Venkata Manohara Reddy A

Indian Institute of Astrophysics, Bangalore

**Abstract:** Matter collapsing to a singularity in a gravitational field is still an intriguing question. Similar situation arises when discussing the very early universe or a universe recollapsing to a singularity. It is suggested that inclusion of mutual gravitational interactions among the collapsing particles can avert a singularity and give finite value for various physical quantities like entropy, density, etc.

It is now generally accepted that the universe went through a very hot dense phase and many of the predictions of this standard evolutionary model, including the anisotropies[1,2] and the light element abundances[3] are now well established.[4] However there are still some enigmatic[5] aspects of the earliest phase of the universe, which are only partially explained by models such as those involving inflation.[6]

In the standard evolutionary big bang cosmological model, the radius of the expanding universe is related to the corresponding temperature of the radiation by:

$$RT = constant \qquad \ldots (1)$$

For the present epoch the temperature is of the order of $3^0$ Kelvin and the scale factor (Hubble radius) of the universe is given as:

$$R = \frac{c}{H_0} \sim 10^{28} cm \qquad \ldots (2)$$

This implies $RT = 3 \times 10^{28} cmK$, which remains constant.

At the era when matter and radiation were of comparable energy densities, which also happens to be around the recombination era ($T \approx 4000K$)

$$R \approx 10^{25} cm \qquad \ldots (3)$$

During the nucleosynthesis era, when the light elements were synthesised, $T \approx 10^{10} K$, $R \approx 3 \times 10^{18} - 10^{19} cm$ and so on.

However, the above relation would imply that in the early universe, when the temperature $T$ was ~Planck temperature $\sim T \sim \left(\frac{\hbar c^5}{G}\right)^{1/2} \frac{1}{k_B} \sim 10^{32} K$, the scale factor $R \sim 10^{-3} cm$.

Thus we see that the universe does not correspond to the Planck dimension at the Planck temperature! This is what we would expect, i.e., at $t_{Pl} \approx 10^{-44} sec$, $R \sim 10^{-33} cm$. So if the universe recollapses, one would expect the same $RT = constant$ relation to hold, but $T \sim 10^{32} K$ corresponds to $R \sim 10^{-3} cm$! Whereas, while it started expansion $R \sim 10^{-33} cm$ which should correspond to a temperature $T \sim 10^{32} K$.

The explanation can be given as follows. In the universe, the total number of the photons $\left(\sim 10^{90}\right)$ is conserved. Another constant is the entropy, i.e. the entropy per baryon. The entropy of the radiation is the total number of photons, i.e. $\sim 10^{90} k_B$ and the entropy per baryon $\sim 5 \times 10^9$.

So if the universe recollapses, when the temperature reaches $T_{Pl}$, each of these photons has an average energy $E_{Pl} \sim 10^{16} ergs (10^{28} eV)$. So that the total energy (due to the blue shifting of each photon energy) of the universe is now $10^{90} E_{Pl}$. So corresponding to the Planck density (i.e. $\sim a T_{Pl}^4$) this implies a radius of just $R \sim 10^{-3} cm$. (If this universe reexpands by a factor of $10^{28}$ or $10^{29}$ the temperature of each photon would drop by a few degrees. I.e. the total energy would be taken up by the redshift of the expansion)

The expansion rate of the universe is given by:
$$\frac{\dot{R}^2}{R^2} = \frac{8\pi G \rho}{3} \quad \ldots (4)$$

Where we can write the density as:
$$\rho = \frac{m}{\tfrac{4}{3}\pi \left(\frac{h}{mc}\right)^3} = \frac{m^4 c^3}{\tfrac{4\pi}{3} h^3}, \quad \ldots (5)$$

as in the early epoch the particles are separated by the Compton length.
The expansion rate is then given by:
$$H^2 = \frac{\dot{R}^2}{R^2} = \frac{2G m^4 c^3}{h^3} = \frac{2 m^2 c^4}{h^2}\left(\frac{Gm^2}{hc}\right) = \frac{2 m^2 c^4}{h^2} \frac{m^2}{hc/G} = \frac{2 m^2 c^4}{h^2} \frac{m^2}{m_{Pl}^2}$$
$$H = \frac{mc^2}{h}\frac{m}{m_{PL}} \Rightarrow H = \frac{m_{Pl} c^2}{h} \text{ for } m = m_{Pl} \quad \ldots (6)$$

However we have to include the mutual gravity between the particles. Then the additional term in the energy density is given by:
$$\rho_G = \frac{Gm^2}{r^4}\Big/c^2 \; ; r = \frac{h}{mc} \quad \ldots (7)$$

So we have:
$$\rho_G = \frac{Gm^6 c^2}{h^4} \quad \ldots (8)$$

This is the binding energy density.
So the expansion rate can be written as:
$$\frac{\dot{R}^2}{R^2} = \frac{8\pi G m^4 c^3}{3 h^3}\left(1 - \frac{Gm^2}{hc}\right) \quad \ldots (9)$$

Including a cosmological constant curvature term we have:

$$\frac{\dot{R}^2}{R^2} = \frac{8\pi G m^4 c^3}{3h^3}\left(1 - \frac{Gm^2}{hc}\right) + \frac{\Lambda c^2}{3} \quad \ldots (10)$$

For $m \approx m_{Pl}$ the term inside the bracket cancels out. This implies that the contraction stops and due to the dark energy term, which at that epoch is ~Planck value, expands exponentially. At $m << m_{Pl}$, this term is negligible. It is important only in the earliest stages when $m \approx m_{Pl}$.

If there were no curvature (cosmological constant) term, the universe would have remained static. As the first two terms in equation (10) cancel at Planck energies $(m \approx m_{Pl})$, the $\Lambda$ term which is ~ $\Lambda_{Pl}$, at that stage, would give rise to an exponential expansion (or re-expansion). The negative pressure violates the strong energy condition, so collapse is not inevitable. However the total entropy remains a constant, i.e. corresponding to the ~ $10^{90} k_B$ photons of the microwave background.

Next such similar situation should also arise for an object collapsing to form a black hole. For the comoving observer who is falling in with the matter, the horizon corresponding to the Schwarzschild radius is nothing special. He is expected to fall into the singularity $r = 0$, which is hidden from the outside (asymptotic) observer by the event horizon (the surface $r = 2m$) which is a one way membrane, according to the cosmic censorship hypothesis.[7]

However by an argument similar to the case of the cosmological collapse discussed above, the observer would not experience infinitely large temperatures, densities or curvatures (which are all expected to blow up at the singularity).

It is well known that inside $r = 2m$, motion can never be static and the collapse of the matter inside (i.e. below $r = 2m$) is described by a metric of the Robertson-Walker type,[8] having the same relation for the scale factor, i.e.,

$$\frac{\dot{R}^2}{R^2} = \left(\frac{8\pi G\rho}{3}\right)^{1/2} \quad \ldots (11)$$

Again taking the mutual self gravity of the interacting particles, separated by a Compton length, the mass (energy) of each of the collapsing particles increases, i.e. it is blue shifted, as in the collapsing cosmic case, so that we have similarly:

$$\frac{\dot{R}^2}{R^2} = \frac{8\pi G m^4 c^3}{3h^3}\left(1 - \frac{Gm^2}{hc}\right) + \frac{\Lambda c^2}{3} \qquad \ldots (12)$$

When $m$ becomes $m_{Pl}$, i.e. the upper limit allowed by quantum gravity, the energies and temperatures is $E_{Pl}$ and $T_{Pl}$, and not infinity as in the classical case, we have the collapse halting ($\dot{R} = 0$), at a finite radius (not zero!) given by:

$$R_{min} \approx N^{1/3} L_{Pl} \qquad \ldots (13)$$

Average separation between particles (each of which are now at the Planck temperature!) is of the order of $L_{Pl}$ and $N$ is the total number (conserved) of particles. For a solar mass, $N \approx 10^{57}$, so that,

$$R_{min} \approx 10^{19} \times 10^{-33} \approx 10^{-14} cm \qquad \ldots (13)$$

(For the universe we had $R_{min} \approx 10^{-3} cm$ at recollapses!)

Here the total entropy is again conserved, i.e. $\approx 10^{57} k_B$.

Note that although the total energy due to the blue shift (during collapse from $r = 2m$ to $r = R_{min}$) is now $E = N k_B T_{Pl}$, the entropy $dE/dT$ is still $Nk_B$, the same as before. When the stellar collapse started, the entropy of the star was $Nk_B$, it continues to remain the same, as far as the comoving observer is concerned(!) even at the horizon and through the fall into the horizon till $R_{min}$ is reached.

At the horizon, for a solar mass star, the temperature of the matter is $\approx 10^{13} K$, but the entropy is still the same for comoving observer. When the collapse reaches $R_{min}$, it again reexpands (driven by the inflationary curvature term $\sim L_{Pl}^{-2}$)

The expansion inside the black hole horizon would again be described by $RT = constant$ (like in the cosmological case, as the dynamics inside is Robertson-Walker as noted above).
Here $RT = constant \sim 10^{19} cmK$ (for solar mass black hole).

So at $R \approx 10^6 cm$ (the horizon radius) the matter inside the black hole stops expanding, has a temperature of $\sim 10^{13} K$ (it has red shifted from $R_{min}$ to $R_{hor}$, which it had when entering the horizon). It now recollapses till $R_{min} \approx 10^{-14} cm$, when the temperature is now $\sim 10^{32} K$ (i.e.

$T_{Pl}$) (*RT = constant!*) and the entropy remains same. It collapses to $R_{min}$ and again reexpands. The outside (asymptotic) observer would not be able to see the collapse or re-expansion inside the horizon.

As far as the comoving observer is concerned, there is no entropy problem. Quantum effects can avert the $r = 0$ singularity. The outside observer however sees a much larger entropy[9] (this is also explained in ref. [10]) as he loses all information about the region $R < 2m$ and all the entropy he estimates is that associated with the horizon, which now behaves like a black body at Hawking temperature given by (for solar mass):

$$T_H = \frac{\hbar c^3}{8\pi GM} \approx 10^{-7} K \qquad \ldots (14)$$

This is nothing but the red shift (by a factor of $10^{19}$) from $10^{13} K$ as measured by a comoving observer at the horizon. This gives a factor of $10^{19}$ increase in entropy which is associated with the horizon, i.e. $E/T$ is the entropy and as $T$ decreases by $10^{19}$ (due to red shift) entropy also increases by the same factor, as seen by the outside observer.[11]

For the comoving observer nothing has changed. For $10^9$ solar mass black hole, $R_{min} \approx 10^{-11} cm$ (scaling as $\left(M/M_{sun}\right)^{1/3}$). The comoving entropy is $\approx 10^{67} k_B$, outside observer would measure $k_B \left(M/m_{Pl}\right)^2$, which is the horizon entropy. Red shift would be $10^{28}$, corresponding to $T_H \sim 10^{-16} K$.

Similar conclusions for a finite minimal radius during collapse (leading to a possible averting of an eventual singularity) can also arise in superstring models where there is a minimal length (set by the string scale). This leads to a generalised uncertainty principle[12] (GUP) and also a modified quantum phase space.[13, 14] The implications are similar, except that $L_{Pl}$ and $m_{Pl}$ in the above relations are replaced by the corresponding string scales $l_{St}$ or $m_{St}$.

Again it turns out that (like the entropy) the total phase space volume in units of $\hbar^3$ is conserved. This can be seen from the fact that as (for example in the case of the collapsing universe or for collapsing matter inside the black hole) the length scale shrinks by $\sim 10^{30}$, the energy (momentum) increases by same factor, so $(d^3p)(d^3x)/\hbar^3$ remains same.

For the universe, this is given by $(d^3x) = V \approx 10^{85} cm^3$; $p^3 \approx (10^{66})^3 = 10^{198}$, so

$$(d^3p)(d^3x)/\hbar^3 \approx 10^{360} \qquad \ldots (15)$$

This is an invariant, as $ER$ is a constant.

Same situation can be verified to occur for a black hole collapse. For solar mass black hole this dimensionless invariant ratio is $\sim 10^{240}$. So both thermodynamics and quantum statistics give rise to quantities which remain invariant.

Again the total action is also invariant. For the universe $\frac{McR}{\hbar} \approx 10^{120}$ is invariant as $MR$ is constant.

In this paper we have discussed some intriguing aspects of the early universe and dynamics of matter inside a collapsing black hole. We have shown that inside a black hole, matter will not collapse to a singularity as expected; causing the break down of known laws of physics; but will only collapse to a minimum radius, if we consider the mutual gravity between the particles.

Based on this argument we can see that for a comoving observer there is no change in entropy as he goes through the horizon. Matter collapses to the minimum radius and re-expands due to the cosmological constant term. Just as the proper mass for a comoving observer is invariant, the entropy also remains invariant for him. The energy of a system is different for different observer; similarly, the entropy is also measured differently by the asymptotic observer. But the comoving observer sees no change in the entropy.

We can also see that the phase space inside a collapsing black hole is also invariant. Hence both thermodynamics and quantum statistics give rise to conserved quantities even inside a collapsing black hole.